\documentclass[aps,twocolumn,amsmath,amssymb,floatfix,showpacs,
groupedaddress]{revtex4}

\usepackage{bm,amsmath,amssymb,amsfonts}

\usepackage[dvips]{graphicx}
\usepackage{array}

\usepackage{color}
\definecolor{brown}{rgb}{0.70,0.30,0.05}
\begin{document}

\title{\textcolor{blue}{Flat band analogues and flux driven extended electronic 
states in a class of geometrically frustrated fractal networks}} 

\author{Atanu Nandy}
\email{atanunandy1989@gmail.com}

\author{Biplab Pal}
\email{biplabpal2008@gmail.com}

\author{Arunava Chakrabarti}
\email{arunava_chakrabarti@yahoo.co.in}

\affiliation{Department of Physics, University of Kalyani, Kalyani,
West Bengal-741235, India}

\begin{abstract}
We demonstrate, by explicit construction, that a single band tight binding 
Hamiltonian defined on a class of deterministic fractals 
of the $b=3N$ Sierpinski type can give rise to 
an infinity of dispersionless, flat-band like states which can be worked out 
analytically using the scale invariance of the underlying lattice. The states 
are localized over clusters of increasing sizes, displaying the existence 
of a multitude of {\it localization areas}. The onset of localization can, 
in principle, be `delayed' in space by an appropriate choice of the energy of the 
electron. A uniform magnetic field threading the elementary plaquettes of the 
network is shown to destroy this {\it staggered localization} and generate 
absolutely continuous sub-bands in the energy spectrum of these non-translationally 
invariant networks.
\end{abstract}


\maketitle
\section{Introduction}
\label{intro}
The interplay of geometry and quantum interference leading to exotic band structure 
in a class of quasi one-dimensional geometrically frustrated lattices  
(GFL) has drawn considerable interest in recent times. The GFL have been 
quite exhaustively studied over the years mainly in the context of localized 
spin systems~\cite{moessner1}, using the Heisenberg model for 
studying frustrated antiferromagnetism~\cite{kikuchi}-\cite{moessner2}, 
or itinerant ferromagnetism within the Hubbard model~\cite{mielke,tasaki}.
The degeneracy induced by the geometric frustration and non-trivial ground 
states such as the spin liquid states as its consequence~\cite{moessner1} have 
unveiled rich physics of the spin systems.

Parallely, electronic spectra of several simple lattices, studied mainly within a 
single band tight binding formalism and only with nearest neighbor hopping, 
have displayed exotic features such as the appearance 
of dispersionless, flat bands (FB)~\cite{lopes}-\cite{leykam}. Such frustrated 
hopping models are important as they offer prospects of strong 
interaction physics such as the fractional quantum Hall effect. 
Graphene anti-dot lattices~\cite{mihajlo} and post-graphene materials 
such as the phosphorene~\cite{ezawa} have been proposed as systems exhibiting 
the qualities for quantum computation~\cite{pedersen} and an electrical tunability 
of quasi-flat bands respectively. 
Recent study on ultracold fermionic atoms patterned into finite 
two dimensional optical lattices of a kagom\'{e} geometry has added to such 
excitements~\cite{chern}. Advancement in controlled growth of artificial, 
tailor-made lattices having atoms trapped in optical lattices 
with the complications of even a 
kagom\'{e} structure~\cite{boong,tamura} has contributed towards increasing 
interest in these deceptively simple looking systems.

\begin{figure}[ht]
\centering
\includegraphics[clip,width=4cm,angle=0]{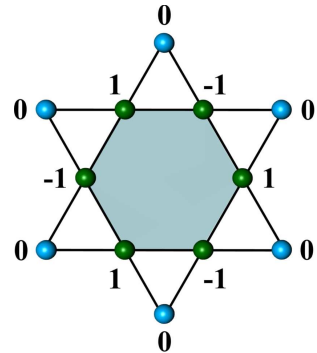}
\caption{(Color online) Schematic view of a part of an 
elementary star-network illustrating the localized distribution 
of amplitudes within a finite cluster.} 
\label{star}
\end{figure}

The non-dispersive character of the 
energy ($E$)-wave vector ($k$) curve makes the {\it effective mass} of the 
electron infinite, making the electron having such an energy 
practically {\it immobile} in the 
lattice. The corresponding single particle state is sharply localized at 
a point, or in a finite cluster of points in the lattice. Such clusters  
are separated from the neighboring clusters of the lattice by vertices where  
the amplitude of the wave function is zero.
The energy eigenvalue corresponding to such a construction belongs to a flat band. 
Fig.~\ref{star} displays one such case, where the central hexagonal cluster 
has non-zero amplitude at the vertices, while the subset of atomic sites residing 
at the vertices of the triangle are zero. This figure corresponds to 
energy $E=-2$ in a tight binding 
description where the `on-site' potential at every lattice point has been 
set equal to zero, and the nearest neighbor hopping integral has been assigned 
the value unity. Such situations can also be realized  
in the kagom\'{e} lattice~\cite{syozi}, or other FB systems~\cite{kubo}. 

The FB systems have been exemplified so far in terms of periodic 
lattice models. The number of flat bands in these systems is always finite. 
Comparatively speaking, practically nothing  
is known about the existence of dispersionless,  
flat band modes in hierarchically generated artificial lattices, where, 
because of the inherent scale invariance, it might be possible to identify
an infinity of such modes.

Here we address this issue in terms of a class 
of Sierpinski gasket (SPG) lattices, which have been widely studied 
over the years~\cite{domany}-\cite{wang} in respect of their 
electronic properties. As one should 
appreciate, the FB states are localized states with finite support in a 
lattice. Though, Anderson localization with weak disorder 
has been addressed for tight binding models with flat bands~\cite{shukla}, 
the effect of self similar lattice topology in this context is 
something which we find practically unaddressed.

This leads to another, 
and a very important additional motivation for the present work,  
namely, an exact evaluation of at least a subsection 
of localized eigenstates in a hierarchical lattice. 
As such lattices are not periodic, one generally has to resort to 
an exact diagonalization of the Hamiltonian to know the energy eigenvalues, and 
because of the fragmented, Cantor set-like 
energy spectrum usually offered by these lattices~\cite{domany} the 
eigenvalues obtained from one finite generation of the lattice in general,  
slip out of the spectrum on increasing the size of the system. A precise  
knowledge of the energy spectrum in the thermodynamic limit thus 
turns out to be a non trivial problem. Though in recent times the existence 
of an infinity of extended, fully transmitting eigenstates have been 
proven within the framework of real space renormalization group (RSRG) schemes
~\cite{wang,arunava}, the exact evaluation of {\it localized state}-eigenvalues 
has still largely remained an unresolved issue, until very 
recently~\cite{biplab,atanu} and, for a special kind of deterministic 
fractal geometries.
\begin{figure}[ht]
\centering
\includegraphics[clip,width=8.5cm,angle=0]{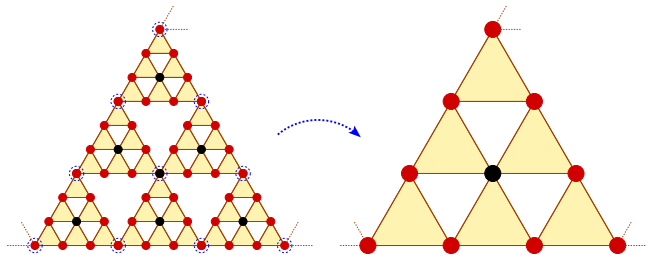}
\caption{(Color online) Portion of an infinite $b=3$ Sierpinski gasket (left) 
and its renormalized version. 
The circles colored red and black have coordination numbers four and six 
respectively. The encircled vertices survive after decimation, leading  
to the scaled lattice (right).} 
\label{decimation}
\end{figure}

In this communication we focus on the so called $b=3N$ Sierpinski gasket 
family~\cite{stanley}. We show that, it is possible to explicitly `construct' 
localized eigenstates spanning finite sized clusters on such lattices. The 
clusters supporting such localized distribution of amplitudes of the wave function 
are {\it separated} from neighboring clusters 
by vertices where the amplitude of the wavefunction is zero at the 
specified energies. As a result, the electron with such 
energies are forced to hop around and remain 
confined in these finite clusters. The minimum size of the clusters 
supporting these localized states increases with the value of $n$. Using 
an RSRG scheme it is possible to unravel an infinity of such localized states and 
the corresponding spanning clusters for every member of the $b=3N$ family. We 
present explicit results for the $b=3$ gasket. The remaining members can be 
easily dealt with using the same formalism, but with increasingly cumbersome 
algebra.

In the second phase of our work we examine the effect of a perturbation 
on the FB states in such hierarchical networks. The perturbation is chosen in the 
form of a uniform magnetic field piercing each elementary plaquette in the $b=3$ 
gasket. The flux threading each plaquette produces 
remarkable changes in the energy spectrum, generating apparent absolute 
continuity in the energy spectrum. Such subbands are populated only 
by extended wavefunctions, as revealed by exhaustive numerical 
study based on the RSRG scheme. 
The fractal character of the energy spectrum is removed, at least locally, 
and the lattice becomes completely transparent to an incoming electron 
when the Fermi level is chosen to lie in such continuous subbands. 
The FB states apparently disappear (at least, they 
could not be explicitly be constructed any more).

In what follows we describe our findings. In section~\ref{sec2} we describe 
the Hamiltonian and the RSRG scheme. Section~\ref{localized} contains a discussion 
of the construction of the localized eigenstates on finite supports. 
A critical discussion on the flat band status of such states are provided in 
Section~\ref{flat}. Section~\ref{magnetic} highlights the effect of 
inserting a uniform magnetic flux in each elementary plaquette 
of the $b=3$ gasket, and the two-terminal transport 
calculations are discussed in Section~\ref{transport}. Finally, in 
section~\ref{conclu} we draw our conclusions. 
\section{The Hamiltonian and the RSRG scheme}
\label{sec2}
We refer to Fig.~\ref{decimation} where a portion of a $b=3$ SPG network (left) 
and its renormalized version are shown (right).  
The electron's hopping is restricted only along each edge, and between nearest 
neighbors only. We distinguish between two types of sites, viz., one with 
coordination number four and colored in red 
(henceforth called $C_4$ sites) and the other, black circles, 
having a coordination number six ($C_6$ sites).
The system is described by the tight binding Hamiltonian,  
\begin{equation}
{\bm H} = \sum_{i} \epsilon_{i} c_{i}^{\dagger} c_{i} + \sum_{\langle ij 
\rangle} t_{ij} \left[c_{i}^{\dagger} c_{j} + h.c. \right] 
\label{hamilton}
\end{equation}
$\epsilon_{i}$ is the on-site potential at each vertex, and will be 
assigned symbols $\epsilon_4$ and $\epsilon_6$ for the $C_4$ and $C_6$ sites  
respectively. $t$ is the nearest neighbor hopping integral.
We shall be interested only in the localization induced by the topology 
of the lattice, and hence, in all numerical results we shall set 
$\epsilon_4=\epsilon_6=0$, and $t=1$. However, on renormalization, the 
on-site potentials deviate from each other at every stage, and 
that is why it is necessary 
to distinguish them at the very beginning.
\begin{figure*}[ht]
\centering
\includegraphics[clip,width=16cm,angle=0]{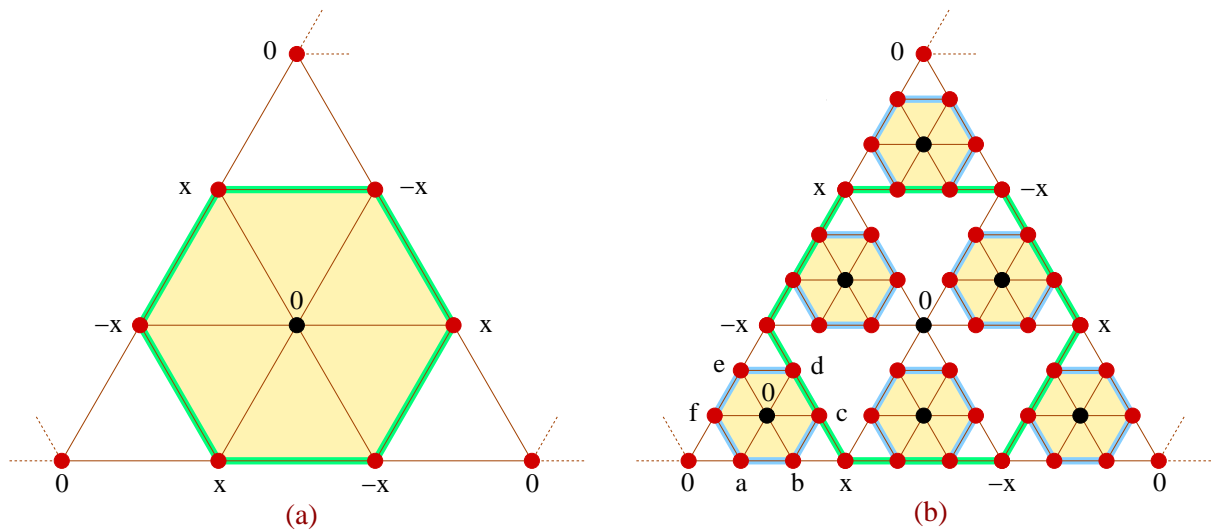}
\caption{(Color online) (a) Distribution of a flat-band state amplitudes 
on a $b=3$ SPG network at its bare scale of length, and (b) the distribution 
on a one step renormalized gasket. The localized state eigenvalues 
are extracted from the equations $E=\epsilon_{4,0}-2t_{0}$ and 
$E=\epsilon_{4,1}-2t_{1}$ respectively.}
\label{amplitude}
\end{figure*}

Using the set of difference equations satisfied by the electrons, viz., 
\begin{equation}
(E-\epsilon_{i}) \psi_{i} = \sum_{j}t_{ij} \psi_{j}
\label{difference}
\end{equation}
we obtain a scaled version of the original lattice, as shown in Fig.~\ref{decimation} . 
An appropriate subset of the $C_4$ and $C_6$ sites 
are decimated to yield an identical geometry with 
the renormalized on-site potentials and the hopping integral. The 
`surviving' sites on the renormalized lattice are encircled in the 
left portion of Fig.~\ref{decimation}. The RSRG recursion 
relations for the on-site potentials and the hopping integrals are given by,
\begin{eqnarray}
\epsilon_{6,n+1} & = & \epsilon_{6,n} + 3 {\cal{F}}_{n} \nonumber \\
\epsilon_{4,n+1} & = & \epsilon_{4,n} + 2 {\cal{F}}_{n} \nonumber \\
t_{n+1} & = & \frac{\beta_{2,n}E^2 + \gamma_{2,n} E + \delta_{2,n}}{\Delta_{n}}
\label{recursion}
\end{eqnarray}
where, ${\cal{F}}_{n}=(\alpha_{1,n}E^3 + \beta_{1,n}E^2 + \gamma_{1,n}E + \delta_{1,n})/
\Delta_{n}$, and, 
\begin{align}
\alpha_{1,n} = &\ 2t_{n}^2 \nonumber \\ 
\beta_{1,n} = &\ -2t_{n}^2 (t_{n}+2 \epsilon_{4,n}+\epsilon_{6,n}) \nonumber \\
\gamma_{1,n} = &\ -2t_{n}^2 [5t_{n}^2-t_n(\epsilon_{4,n}+\epsilon_{6,n})- \nonumber \\
&\ \epsilon_{4,n}(\epsilon_{4,n}+2\epsilon_{6,n})] \nonumber \\
\delta_{1,n} = &\ -2t_n^2 [2t_n^3 + 
t_{n}\epsilon_{4,n}\epsilon_{6,n} + 
\epsilon_{4,n}^2\epsilon_{6,n} - \nonumber \\
&\ t_{n}^2 (4 \epsilon_{4,n} + \epsilon_{6,n})] 
\label{symbols1}
\end{align}
The other symbols used 
are, 
\begin{align}
&\beta_{2,n} = t_{n}^3 \nonumber \\
&\gamma_{2,n} = t_{n}^3 (4t_n-\epsilon_{4,n}-\epsilon_{6,n}) \nonumber \\ 
&\delta_{2,n} = t_{n}^3 (\epsilon_{4,n}\epsilon_{6,n} + 2t_{n}^2 - 4t_n\epsilon_{4,n}) 
\nonumber \\ 
&\Delta_{n} = [(E-\epsilon_{4,n})^2 - t_{n}^2][(E-\epsilon_{6,n})(E-\epsilon_{4,n}-2t_{n})-6t_{n}^2]
\label{symbols2}
\end{align}
The `stage' of renormalization is denoted by $n \ge 0$, and we have chosen to stick to 
$\epsilon_{4,0}=\epsilon_{6,0}=\epsilon$ and $t_0=t$ at the bare scale of length.

The recursion relations Eq.~\eqref{recursion} will now be exploited to 
construct and analyze a class of localized and extended states in such systems.
\section{Constructing the localized states}
\label{localized}
\subsection{The basic scheme}
Let us begin with the $b=3$ gasket and refer to Fig.~\ref{amplitude}. 
It is simple to work out that, if we 
set $E=\epsilon_{4}-2t$, a consistent solution of the difference equations (or 
equivalently, the Schr\"{o}dinger equation) is obtained in which the amplitude 
of the wave function $\psi_{j}$ is zero at the $C_{6}$ vertices with coordination 
number six and lying at the center of the shaded hexagon, 
as well as on a subset of $C_{4}$ vertices. The non-zero amplitudes are 
designated by $x$ (which can be set equal to unity, for example). 
This is illustrated in the left panel of Fig.~\ref{amplitude}.
This latter subset of $C_{4}$ sites sit at the `connecting' points 
between hexagonal plaquettes at different portions of the infinite 
lattice. By making the amplitude vanish at these strategic sites it 
is possible to {\it confine} the hopping of the electron around the 
periphery of the shaded hexagons. The eigenstate resembles the spirit 
of a molecular state~\cite{kirkpatrick} localized partially 
by quantum interference and partially by physical boundary formed by the sites 
with zero amplitude. The wave function corresponding 
to this energy eigenvalue does not have any overlap with the wave function 
described on the next neighboring hexagons. 
The localization {\it area} is the shaded 
hexagonal plaquette confining the 
electron's motion for $E=\epsilon_{4}-2t$ along the `periphery', and is
distributed throughout the infinite geometry.
\subsection{Exploiting the self similarity}
Once such a construction of localized state-amplitudes is made possible in the 
bare length scale, the self similar structure of the SPG lattices can be 
exploited to construct wave functions, and to `work out' the corresponding 
areas of localization at various scales of length. The eigenvalues 
corresponding to such cases are easily extracted from the solutions of the 
equation 
\begin{equation}
E = \epsilon_{4,n} - 2t_{n}
\label{root}
\end{equation}
at any desired $n$-th stage of RSRG. For every real root of this 
equation, a construction similar to that in the original lattice 
can be made in respect of the values of the amplitudes of the 
wave function at different $C_{4}$ and $C_{6}$ vertices on the 
scaled version of the lattice. One then has to work in the 
reverse direction to lay out the amplitude-distribution on the 
original lattice at the bare scale of length by simultaneously solving 
the difference equations Eq.~\eqref{difference}
for the respective energy eigenvalue.
The scheme is illustrated in Fig.~\ref{amplitude}(b). 
With $\epsilon_{4}=\epsilon_{6}=0$ and 
$t=1$, we get $E=-2$ at the original length scale, while the same parameters yield 
localized states at $E=-2$ and $E=1 \pm \sqrt{2}$ when we set
$n=1$, that is, when we solve the equation  
$E=\epsilon_{4,1}-2t_1$. Let us now explain the construction of localized states 
for the {\it new} roots, viz., $E=1 \pm \sqrt{2}$.
  
The hexagon in Fig.~\ref{amplitude}(b) whose sides are highlighted 
by thick green line is the `smallest' 
hexagon on a one step renormalized gasket. 
We choose the distribution of amplitudes 
at $E=\epsilon_{4,1}-2t_1$ around this hexagon exactly in the 
same way as we did for $E=-2$ in the bare length scale (Fig.~\ref{amplitude}(a)).
Such a construction, as we now appreciate, can be made consistently throughout 
the infinite system. The next job is to work out, in the reverse 
direction, the amplitudes of the wave function 
at all the vertices of the original gasket with the prefixed amplitudes 
(with values $\pm x$ at the edges of the green highlighted hexagon and zero at its 
center) for energy eigenvalues $E=1 \pm \sqrt{2}$. 

The amplitudes are denoted by the letters $a$-$f$ in Fig.~\ref{amplitude}(b). 
We have shown explicitly the smaller hexagons (blue highlighted) 
that are embedded in the larger one (green highlighted).
It is possible to obtain a consistent set of values for the 
amplitudes at all vertices of the smaller (blue) hexagons. 
For example, with $x=1$, set arbitrarily, 
the distribution of amplitudes for $E=1+\sqrt{2}$ (one solution of 
$E=\epsilon_{4,1}-2t_1$ with $\epsilon=0$ and $t=1$) 
is given by, $a=-f=(\sqrt{2}-1)/2$, $b=-e=1/\sqrt{2}$ and 
$c=-d=1/2$. The amplitude at the center of a small hexagon is zero for this 
particular distribution. An identical pattern of distribution, but with 
a different set of values for $a$, $b$, $\hdots\ f$, can be easily obtained 
if one sets $E=1-\sqrt{2}$.
\begin{figure*}[ht]
\centering
\includegraphics[clip,width=16cm,angle=0]{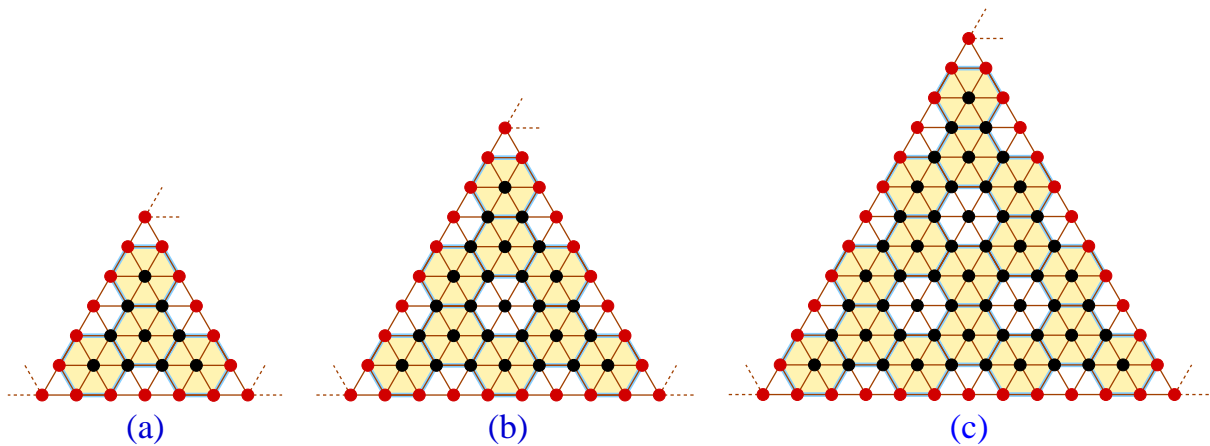}
\caption{(Color online) Distribution of amplitudes for 
a non-dispersive eigenstate at $E=\epsilon_{4}-2 t$ 
on $b=6$, $b=9$ and $b=12$ Sierpinski gaskets. With increasing scale 
factor, the amplitudes assume non-trivial distribution over larger and 
larger clusters which ultimately get decoupled from the rest of the lattice, 
courtesy, the zero amplitudes at the connecting vertices.} 
\label{staggered}
\end{figure*}

Needless to say, the scheme can be extrapolated to unravel other localized states 
by solving the equation $E=\epsilon_{4,n}-2t_{n}$ at any desired level. The roots 
increase in number, and the set of solutions obtained at the $n$-th stage 
of renormalization contains all the roots obtained from the $(n-1)$-th stage.
In every case one has to identify the smallest hexagon in that scale of length, 
set the amplitude of the wave function equal to {\it zero} at its center. The 
sites decorating the periphery of the basic hexagon then should be assigned 
amplitudes equal to $\pm x$ as we did in Fig.~\ref{amplitude}. 
Such a basic hexagon at the renormalized scale must be {\it separated} from 
its neighboring hexagons, a condition that is easily satisfied by setting 
the amplitude at the `connecting' sites equal to zero.
The final task is to map the distribution back onto the original 
lattice and to solve the set of equations Eq.~\eqref{difference} 
to work out the amplitudes at the inner sites of the 
lattice at its bare scale of length. 
\subsection{The `staggered' character of localization}
An interesting qualitative distinction between the energy eigenvalues 
corresponding to these localized states can be unravelled if we look at the 
evolution of the hopping integral under RSRG steps. For any energy eigenvalue 
obtained from the $n$-th stage, and 
by solving the equation $E=\epsilon_{4,n}-2t_n$, 
the hopping integral remains finite upto 
$n$-th RSRG steps. This means that one has non-zero overlap of the wave function 
between nearest neighboring sites at that length scale. When mapped back 
onto the original lattice it amounts to a non-zero overlap of the wave function 
between sites much beyond the nearest neighbors. However, the hopping integral 
starts decaying after the $n$-th RSRG step implying that the wave function 
looks `extended' only {\it locally}, and in reality such wave functions are 
localized on a finite cluster of the infinite lattice. 
The overlap of the envelopes of the wave function for a localized state-energy 
extracted at the $n$-th stage weakens beyond a certain scale of length.
Such finite sized clusters span larger and larger areas in real 
space as one extracts the energy eigenvalue by solving the above 
equation at deeper and deeper scales of length.Thus simply 
by `choosing' the value of stage of renormalization '$n$' one 
can {\it delay}, in space, the localizing character of the wave 
function. In a recent study~\cite{biplab} we encountered such 
{\it staggered localization} in a special class of fractal lattices.

Before we end this section, it is pertinent to point out that the size 
of the spanning clusters even at the bare length scale depends on 
the value of $n$ in the class of $b=3N$ SPG fractals.
Fig.~\ref{staggered} depicts three other cases, viz., $b=6$, $b=9$ and 
$b=12$. We stick to the same value of energy, viz., $E=\epsilon_{4}-2t$.  
The localization areas even at the bare length scale are already larger than 
that in the $b=3$ case. In each case, the electron's hopping 
is confined around the peripheries of clusters of finite 
size, shown by the shaded areas. Such finite sized clusters 
grow bigger and bigger as one goes from $b=6$ to $b=12$, and 
beyond. The clusters, however big, are ultimately separated from 
each other by a subset of $C_4$ sites with $\psi_{j}=0$, and are 
distributed throughout the infinite geometry.  
\section{The flat band analogues}
\label{flat}
A pertinent question is, whether these infinite number of localized eigenstates 
are non-dispersive in character, in the spirit of the flat band states 
observed in a class of periodic lattices as mentioned in the introduction. 
\begin{figure*}[ht]
\centering
\includegraphics[clip,width=14cm,angle=0]{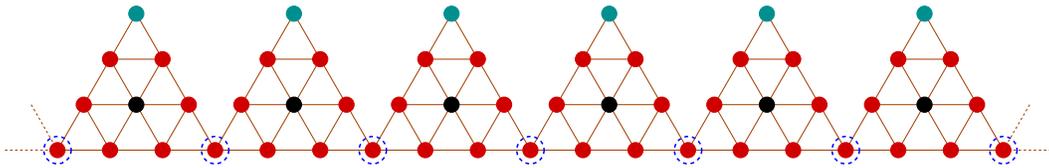}
\caption{(Color online) 
A periodic array of finite generation $b=3$ SPG networks. With 
increasing generation each unit cell will be representative of an infinite 
SPG network.} 
\label{array}
\end{figure*}
The absence of translational invariance in the SPG class of lattices 
prevents a derivation of the usual energy-wave vector dispersion relation.
At the same time, the fact that the localized states discussed so far 
have been `constructed' so as to display a kind of {\it self localization}, 
where the amplitudes are distributed over a finite cluster of lattice points, 
effectively `decoupled' from the rest of the system. This inspires us to undertake 
a couple of indirect, but effective ways of discerning the actual character of 
these states.
\subsection{Dispersion relation for a periodic array of fractal approximants}
First, we construct a periodic array of $b=3$ SPG networks. Each `unit cell' 
consists of a finite SPG at its $n$-th generation. 
One such array comprising the first generation $b=3$ SPG network 
is shown in Fig.~\ref{array} for illustration. As is evident, such a construction 
requires the presence of a top vertex with coordination number two (colored cyan 
in Fig.~\ref{array}). It is important to appreciate that for such finite 
systems the distribution of amplitudes does not follow the method prescribed 
for Fig.~\ref{amplitude}, which was typical of an infinite system where the 
`end points' were not `visible'.

As we gradually increase 
the value of $n$, the unit cell itself turns out to be a reasonably good 
approximant of the infinite system, and becomes the true 
representative in the limit $n \rightarrow \infty$. 
For a periodic array of $n$-th generation SPG networks one can renormalize each 
`unit cell' to $n$-times to convert the array 
into an array of triangles with the potentials at the 
vertices equal to $C_{2,n}$ and $C_{4,n}$ respectively. The 
nearest neighbor hopping integral along the edges of such 
effective triangles are $t_n$. This array of triangles 
is then converted into an array of single `effective' 
atomic sites (encircled by dotted lines in Fig.~\ref{array}) 
with energy dependent potential, and inter-connected by an 
energy dependent hopping integral. 
The effective, energy dependent on-site potential 
and the nearest neighbor hopping integral in the linear chain 
of {\it renormalized} atoms are given by,
\begin{eqnarray}
\tilde\epsilon & = & \epsilon_{4,n} +  \frac{2t_{n}^{2}}{E - \epsilon_{2,n}} \nonumber \\
\tilde t & = & \frac{t_{n}^{2}}{E - \epsilon_{2,n}}
\end{eqnarray}
In this case, we need an additional renormalization of the top vertex with 
coordination number two, viz., 
\begin{equation}
\epsilon_{2,n+1} = \epsilon_{2,n} + {\cal{F}}_{n} \nonumber \\
\end{equation} 
The definition of ${\cal{F}}_{n}$ is already given in earlier section 
(Eq.\eqref{symbols1} and Eq.\eqref{symbols2}).
The $E\text{-}k$ dispersion relation for such 
a periodic chain is then easily obtained through the 
equation, 
\begin{equation}
E = \tilde\epsilon + 2 \tilde t \cos{ka}
\end{equation}

\begin{figure}[ht]
\centering
\includegraphics[clip,width=8.5cm,angle=0]{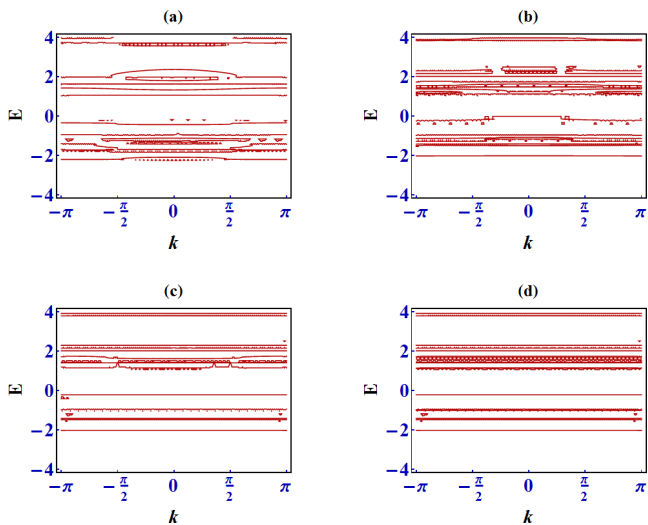}
\caption{(Color online) The dispersion relation for the 
array depicted in Fig.~\ref{array}. The flat, dispersionless modes 
are apparent. The flat band at $E=-2$, the 
root of the equation $E=\epsilon_{4} - 2t$ with $\epsilon_{4}=\epsilon_{6}=0$ 
and $t=1$ has already been shown to exist, by 
explicit construction, for an infinite system.}
\label{dispersion}
\end{figure}

In Fig.~\ref{dispersion} 
we depict the 
dispersion relation when the unit cell consists of (a) the first generation, 
(b) the second generation, (c) the third generation and (d) the fourth generation 
$b=3$ SPG networks. We have set $\epsilon_{4}=\epsilon_{6}=0$ and $t=1$. It is 
clearly seen form Fig.~\ref{dispersion}(c) and (d) 
that flat, non-dispersive bands appear at $E=-2$ and at 
$E=1 \pm \sqrt{2}$ -- the three roots obtained from the equation 
$E=\epsilon_{4,n}-2t_n$ at the bare length scale and from the one step 
renormalized lattice, as discussed in the last section. 
Even with a modest unit cell structure for the triangular array, the 
{\it self (staggered) localized} states corresponding to the truly 
infinite system are showing up.
Thus, we feel tempted 
to identify these self-localized states as flat band analogues of similar 
states in quasi one dimensional systems~\cite{lopes}-\cite{flach}. However, 
we are now in a position to appreciate that such states will be infinite in 
number as the size of the system approaches the thermodynamic limit.
\subsection{ A study of the density of states}
As a second check, we evaluate the local density of states (LDOS) at 
the $C_{4}$ and $C_{6}$ sites. The results are shown in Fig.~\ref{ldos}(a) 
and (b) respectively. This is done following the standard 
procedure, and using the RSRG recursion relations Eq.~\eqref{recursion}, 
which finally yield the local Green's function (LGF) at the $C_{4}$ and 
$C_{6}$ sites as, $G_{00}^{(4)}=(E-\epsilon_4^*)^{-1}$ 
and $G_{00}^{(6)}=(E-\epsilon_6^*)^{-1}$ respectively. The symbol 
$\epsilon_{j}^{*}$ indicates the fixed point value of the respective 
parameter. Finally, the LDOS at the desired site is obtained as, 
$\rho_{j}=(-1/\pi) \lim_{\eta \rightarrow 0}\text{Im}\;G_{00}^{(j)}(E+i\eta)$, with 
$j=4$ or $6$ depending on the coordination number.

The LDOS (in the absence of an external magnetic field) exhibits a highly 
fragmented distribution as seen in Fig.~\ref{ldos}(a) and (b). This is expected. 
The fine structure is related to the distribution of spectral 
weight~\cite{nicolic}, and is controlled by the lattice topology. 
One should observe that the $j$-th site ($C_4$ or $C_6$) having 
$\psi_j=0$ can not contribute to the LDOS at that particular site.
Nevertheless, non-trivial contribution comes from the neighboring 
sites yielding a finite value of the density of states at a $C_{4}$ site.

It is important to appreciate the role of the numerical value of the 
imaginary part $\eta$ added to the energy. $\eta$ determines the width 
of the LDOS around an energy. We have carefully examined the growth 
of the LDOS for  $E=-2$, $1 \pm \sqrt{2}$ and at several other 
localized {\it flat band} states as obtained by explicit construction 
at bare length scale and from a one-step renormalized lattice. 
The result is tabulated in Table~\ref{table} for the above three energy 
eigenvalues. It is explicitly seen that, for each of these energy eigenvalues, 
with a sequential decrease in the value of $\eta$, the LDOS at the 
$C_4$ site, viz., $\rho_4$ increases exponentially, 
indicating that in the limit $\eta \rightarrow 0$, the LDOS diverges 
at these special energy values. 
\begin{table} 
\renewcommand{\arraystretch}{1.2}
\begin{tabular}
{|>{\centering\arraybackslash}m{0.5in}|>{\centering\arraybackslash}m{0.5in}|
>{\centering\arraybackslash}m{0.5in}|>{\centering\arraybackslash}m{0.8in}|}
\hline
{\boldmath\(E\)} & {\boldmath\(\eta\)} 
& {\boldmath\(\rho_{4}(E)\)} & {\boldmath\(\rho_{6}(E)\)}\\
\hline
     & $10^{-2}$ & $3.28$ & $0.03$ \\ 
     & $10^{-3}$ & $15.66$ & $0.07$ \\
     & $10^{-4}$ & $74.38$ & $1.5\times10^{-3}$ \\
$-2$ & $10^{-5}$ & $355$ & $3.2\times10^{-4}$ \\
     & $10^{-6}$ & $1703.2$ & $6.8\times10^{-5}$ \\
     & $10^{-7}$ & $8701$ & $1.34\times10^{-5}$ \\
     & $10^{-8}$ & $40419$ & $2\times10^{-6}$ \\ \hline 
             & $10^{-2}$ & $1.68$ & $1.06$ \\ 
             & $10^{-3}$ & $5.2$ & $0.5$ \\
             & $10^{-4}$ & $22.5$ & $0.11$ \\
$1-\sqrt{2}$ & $10^{-5}$ & $106.4$ & $2\times10^{-2}$ \\
             & $10^{-6}$ & $506.24$ & $5.4\times10^{-3}$ \\
             & $10^{-7}$ & $2402.8$ & $1.13\times10^{-3}$ \\
             & $10^{-8}$ & $11454$ & $2.4\times10^{-4}$ \\ \hline 
             & $10^{-2}$ & $1$ & $7.5\times10^{-2}$ \\ 
             & $10^{-3}$ & $4.41$ & $1.75\times10^{-2}$ \\
             & $10^{-4}$ & $20.91$ & $3.8\times10^{-3}$ \\
$1+\sqrt{2}$ & $10^{-5}$ & $100$ & $8\times10^{-4}$ \\
             & $10^{-6}$ & $473$ & $1.7\times10^{-4}$ \\
             & $10^{-7}$ & $2249$ & $3.6\times10^{-5}$ \\
             & $10^{-8}$ & $10747$ & $7.4\times10^{-6}$\\ \hline    
\end{tabular}
\caption{The values of the local densities of states 
$\rho_4(E)$ and $\rho_6(E)$ at special 
energy eigenvalues $E=-2$, and $1\pm \sqrt{2}$ as obtained from the 
original lattice and its one-step renormalized version 
corresponding to localized flat-band states.
We have set the on-site potential $\epsilon=0$ at each site, and the 
energy is measured in unit of the hopping integral $t$.}
\label{table}
\end{table}
As we know, the average density of states 
(ADOS) is related to the group velocity $v_{g}$ of the electron by the relation 
\begin{equation}
\rho_{av} \propto \int v_{g}^{\:-1} d^{3}k
\end{equation}
where, $v_{g} \propto dE/dk$. Therefore, a diverging LDOS, 
leading to a diverging ADOS, is indicative of the 
zero (or extremely low) group velocity and hence, a {\it non-dispersive} character 
of the state. The energy eigenvalues, obtained by construction, and 
corresponding to the self-localized eigenfunctions in the present example 
can thus be considered as the so called {\it flat band} states. We can thus 
unravel infinite number of such non-dispersive modes in an infinite 
$b=3N$ SPG network.
\section{Effect of a uniform magnetic field}
\label{magnetic}
As is well known, the FBS are unstable against perturbation. We 
examine the general character of the energy spectrum of a 
$b=3$ SPG network when each elementary plaquette is threaded by a 
uniform magnetic flux $\Phi$. The presence of a magnetic field breaks the 
time reversal symmetry of the electron-hopping along a bond, and 
this is taken care of by introducing a Peierls' phase in the hopping 
integral, viz., $ t \rightarrow t \exp{(i2\pi\Phi/3\Phi_0)}$, where 
$\Phi_{0}=hc/e$ is the fundamental flux quantum.

\begin{figure}[ht]
\centering
\includegraphics[clip,width=8.5cm,angle=0]{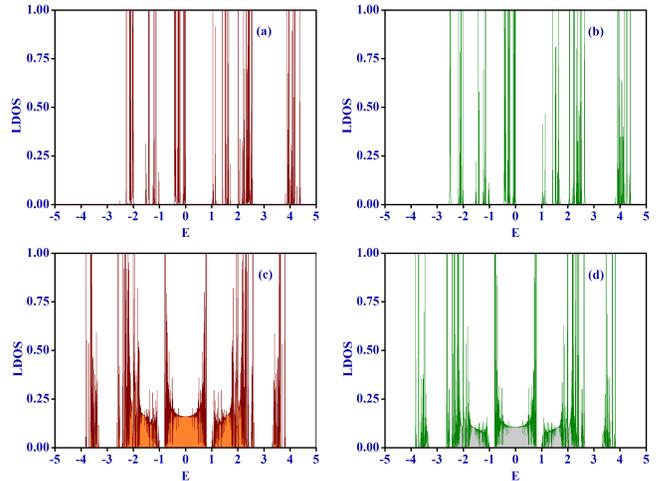}
\caption{(Color online) Local density of states at the (a) $C_{4}$
and (b) $C_{6}$ sites of an infinite $b=3$ SPG network in the absence 
of a magnetic field. (c) Local density of states at a $C_{4}$ site 
when each elementary plaquette is pierced by a magnetic flux $\Phi=\Phi_{0}/4$, 
and (d) the same at a $C_{6}$ site with $\Phi=\Phi_{0}/4$. We have set 
$\epsilon_{4}=\epsilon_{6}=0$ and $t=1$. The imaginary part added to the 
energy is $\eta=10^{-4}$.}  
\label{ldos}
\end{figure}

The RSRG method is used again to evaluate the LDOS, which is shown in 
Fig.~\ref{ldos}(c) and (d) for the $C_{4}$ and $C_{6}$ sites respectively for 
$\Phi=\Phi_{0}/4$. We draw the attention of the reader to the {\it absolutely 
continuous} shaded portions in the LDOS. This is a generic character for all values 
of the piercing flux. We have carefully scanned the continuous subbands in 
arbitrarily small energy intervals, and observed the flow of the hopping 
integral under the RSRG iterations. Extensive numerical search reveals that 
for any energy belonging to the continua, the hopping integral keeps on oscillating 
for an indefinite number of iteration loops. This implies that, for any energy 
forming the continua there is non-zero overlap between the envelopes of the wave 
function at various scales of length, and that, the subbands are 
densely populated by {\it extended} wave functions. In addition, the energy $E=0$ 
lying at the center of the spectrum exhibits a {\it one cycle fixed point} 
in the parameter space, that is, $(\epsilon_{4,n},\epsilon_{6,n},|t_n|) 
= (\epsilon_{4,n+1},\epsilon_{6,n+1},|t_{n+1}|)$, for $n \ge 0$.

\begin{figure}[ht]
\centering
\includegraphics[clip,width=8cm,angle=0]{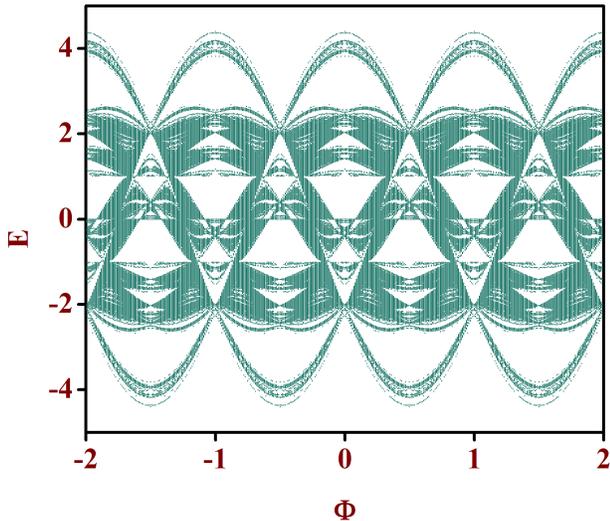}
\caption{(Color online) Distribution of energy eigenvalues against 
the magnetic flux $\Phi$ (in unit of $\Phi_0=hc/e$) in an infinite $b=3$ SPG network.}
\label{spectrum}
\end{figure}

The existence of the subbands of extended states does not rule out 
the other localized states, though the non-dispersive, flat bands 
apparently do not exist (or, at least, become undetectable). As, for an 
electron circulating around a closed loop, the magnetic flux plays a 
role equivalent to the wave vector~\cite{gefen}, we illustrate the 
$E\text{-}\Phi$ diagram (Fig.~\ref{spectrum}) as an analogue of 
the dispersion relation in Fig.~\ref{dispersion}. 
The energy eigenvalues of the fractal network are seen to form mini bands as 
a function of the flux $\Phi$ with period $\Phi_{0}$. 
There are multiple, inter-twined band crossings, and quite a dense 
distribution of eigenvalues, forming {\it almost continuous} 
quasi-flat $E\text{-}\Phi$ `bands'. The pattern is flux periodic 
with a period equal to one flux quantum, and close observation 
of the $E\text{-}\Phi$ `band' subsections 
reveals formation of interesting variants of the Hofstadter 
butterflies~\cite{hofs}. The spectral landscape is a quantum fractal, 
and encoding the gaps with appropriate topological quantum numbers 
remains an open problem for such deterministic fractals.
\section{Two-terminal transmission characteristics}
\label{transport}
For the sake of completeness of the above discussion, 
we have computed the two-terminal transmission characteristics 
of a finite system. The procedure is standard and is often 
used to evaluate the transmission coefficient of such hierarchical fractal 
structures~\cite{biplab}. The basic idea is to clamp the system in between 
two semi-infinite periodic leads, the so called `source' and the `drain'. 
The on-site potential at the atomic sites in the leads is ${\mathcal{U}}_{L}$. 
The hopping integral is ${\mathcal{T}}_{L}$ in between the 
nearest neighboring atomic sites 
in the leads. The finite sized network sandwiched in between the 
two ordered leads is then successively renormalized to reduce it to an 
effective diatomic molecule (dimer)~\cite{biplab}. The transmission coefficient of the 
lead-network-lead system then is given by a well-known formula~\cite{liu},
\begin{eqnarray}
&T=\dfrac{4\sin^{2}ka}{|\mathcal{A}|^{2}+|\mathcal{B}|^{2}} \\
&\text{with,}\quad \mathcal{A}=[(M_{12}-M_{21})+(M_{11}-M_{22})\cos ka] \nonumber\\
&\text{and}\quad \mathcal{B}=[(M_{11}+M_{22})\sin ka]\nonumber
\end{eqnarray}
where, $M_{ij}$ refer to the dimer-matrix elements, written appropriately 
in terms of the on-site potentials of the final renormalized {\it left} (L) 
and {\it right} (R) atoms at the extremities of the finite 
SPG network  
and the renormalized hopping between them. 
$\cos ka=(E-{\mathcal{U}}_{L})/2{\mathcal{T}}_{L}$, 
and $a$ is the lattice constant in the leads which is set equal to unity 
throughout the calculation.

\begin{figure}[ht]
\centering
\includegraphics[clip,width=6cm,angle=0]{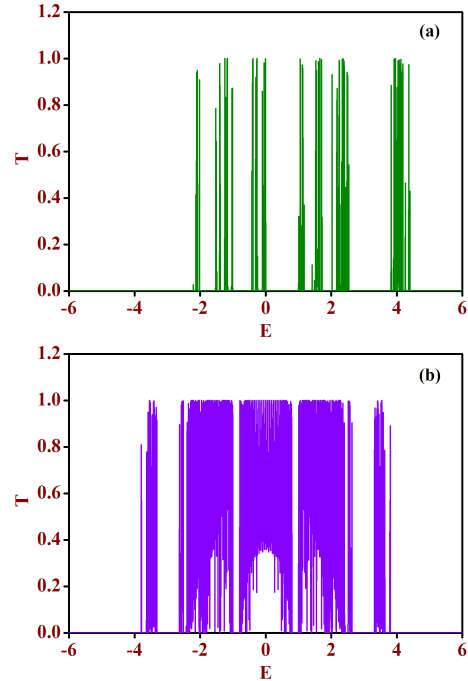}
\caption{(Color online) Transmission coefficient of a $4$-th generation 
$b=3$ SPG network with (a) $\Phi=0$, and (b) $\Phi=\Phi_{0}/4$. We have 
chosen $\epsilon_{4}=\epsilon_{6}=0$, $t=1$, 
$\mathcal{U}_{L}=0$ and $\mathcal{T}_{L}=3$.}
\label{transmission}
\end{figure}

It is apparent from Fig.~\ref{transmission} that, 
the low transmitting character of the gasket in the 
absence of an external magnetic field 
(Fig.~\ref{transmission}(a)) is enriched by high transmitting windows in the 
presence of a magnetic flux (Fig.~\ref{transmission}(b)). 
This corroborates the argument about the 
existence of absolutely continuous subbands populated by extended eigenfunctions.

The transmission coefficient, as expected, is flux periodic, with period 
equal to one flux quantum. The fine structures of the Aharonov-Bohm oscillations 
in transport are of course, sensitive to the chosen energy eigenvalue. However, 
to save space, we refrain from showing these figures.
\section{Conclusion}
\label{conclu}
In conclusion, we have 
unravelled an infinity of flat band-like, dispersionless 
eigenstates in the $b=3N$ class of Sierpinski gasket fractals. 
The states are localized in clusters of finite size, which 
are {\it effectively} separated 
from similar clusters in the lattice by atomic sites 
where the amplitude of the wave function is zero. The 
topology of the lattice confines the dynamics 
of the electron in such local clusters for these special sets of energy 
eigenvalues. The states have been obtained by explicit construction and the 
inherent self similarity of the lattice is exploited through a real space 
rescaling  of the system. The dispersionless character of the flat band states are 
confirmed by obtaining the $E\text{-}k$ dispersion curves of a periodic array 
of the approximants of the true gasket, and also from the diverging 
character of the local density of states at those special energy eigenvalues.
The two terminal transport and the magnetic field induced absolutely 
continuous subbands have been discussed in details. The intricate mixture of 
complexity, frustration and a deterministic geometry may throw up new challenges 
in the quantum Hall physics in such fractal geometries.
\begin{acknowledgements}
AN acknowledges a research fellowship from UGC, India and BP is 
thankful to DST, India for an INSPIRE fellowship. AC thanks Bibhas Bhattacharyya 
and Santanu K. Maiti for enlightening discussions.
\end{acknowledgements}

\end{document}